\documentstyle[12pt,aasms4]{article}
\def\kms{kms$^{-1}$}

\begin{document}

\title{On the use of scaling relations for the Tolman test}

\author{Mariano Moles$^{1,2}$, Ana Campos$^{1,2}$, Per Kj\ae rgaard$^3$,
Giovanni Fasano$^{4}$, and Daniela Bettoni$^4$}

\altaffiltext{1}{Instituto de Matem\'aticas y F\'{\i}sica Fundamental,
CSIC, C/ Serrano 123, 28006 Madrid, Spain}

\altaffiltext{2}{Observatorio Astron\'omico Nacional, Apdo. 1143, 28800
Alcal\'a de  Henares, Madrid, Spain. E-mail: moles@oan.es, ana@oan.es}

\altaffiltext{3}{Copenhagen University Observatory, The Niels Bohr Institute for
Astronomy, Physics and Geophysics, Juliane Maries Vej 30, DK 2100 Copenhagen.
E-mail:per@ursa.astro.ku.dk}

\altaffiltext{4}{Osservatorio Astronomico di Padova, vicolo dell'Osservatorio 5,
35122, Padova, Italy. E-mail:gfasano@leda.pd.astro.it, dbettoni@leda.pd.astro.it}

\begin{abstract}

The use of relations between structural parameters of early type galaxies to perform
the Tolman test for the expansion of the Universe is reconsidered. Scaling relations
such as the Fundamental Plane or the Kormendy relation, require the transformation
from angular to metric sizes, to compare the relation at different z values. This
transformation depends on the assumed world model: galaxies of a given angular size,
at a given z, are larger (in kpc) in a non-expanding universe than in an expanding one.
Furthermore, the luminosities of galaxies are expected to evolve with z in an expanding
model. These effects are shown to conspire to reduce the difference between the
predicted Surface Brightness change with redshift in the expanding and non expanding
cases.

We have considered expanding models with passive luminosity evolution. We find that their
predictions for the visible photometric bands are very similar to those of the static model
till z$\sim$1, and therefore, the test cannot distinguish between the two world models. Recent
good quality data on the Kormendy relation and the Fundamental Plane at intermediate redshifts
are consistent with the predictions from both models. In the K-band, where the expected
(model) luminosity
evolutionary corrections are smaller, the differences between the expanding and static
models amount to $\sim$0.4 (0.8) magnitudes at z = 0.4 (1). It is shown that, due to that
small difference between the predictions in the covered z-range, and to the paucity and
uncertainties of the relevant SB photometry, the existing K-band data is not adequate
to distinguish between the different world metrics, and cannot be yet used to discard
the static case. It is pointed out that the scaling relations could still be
used to rule out  the non-evolving case if it could be shown that the coefficients
change with the redshift.

\end{abstract}

{\sl Subject headings:} galaxies: fundamental parameters --- cosmology: theory ---
cosmology: observations

\section{Introduction}

The Big Bang cosmological model rests upon the Cosmological Principle, which determines the
form of the metric of the world, and allows the explanation of the redshift phenomenon
as a pure geometrical effect directly related to the nature of that metric. The
Microwave Background Radiation or the abundance of the light elements strongly support the
standard model. Thus, given the overall consistency of the model, some authors consider
as superfluous any attempt to test the basic hypothesis, i.e., the form of the metric.
Furthermore, its explanatory capacity is often used to rule out any other theoretical
alternative. Recently, van Dokkum and Franx (1996) have argued that the microwave background
radiation would not be planckian if the Surface Brightness (SB) dimming were different of
$(1+z)^{-4}$. We remark that this is true from within the standard model, otherwise it
should have been abandoned long ago. But the fact that data can be accommodated within the
standard model, important as it is, cannot be considered as a formal proof of the
underlying hypothesis. It is worth to remember that, formally, a false statement can have
consequences that are true.

Direct tests have been proposed since long (see Hubble \& Tolman 1935), and recently
reformulated from different angles (Moles 1991; Sandage \& Perelmuter 1991;
Kj\ae rgaard, J\o rgensen \& Moles 1993). Among the tests proposed by Hubble \& Tolman,
the SB test has been considered
the most powerful and {\it clean}. Indeed, the SB of a standard candle only depends on
the red-shift: it would decrease as $(1+z)^{-4}$ in an expanding Universe. Otherwise,
i.e., in a static, non-expanding universe, with some (unknown) operating mechanism to produce
the observed redshift, the SB dimming would go like $(1+z)^{-1}$. The test
cannot be considered as able to select one among different models since there are no known
alternatives to the standard cosmology based on a non-expanding metric. Here, as stated by
Hubble \& Tolman, the non-expanding case is taken to gauge the standard model.

In practice, when the Fundamental Plane (FP) or Kormendy relations are used, it is necessary
to use the metric to transform the angles we measure into linear sizes, to compare
data at different z values. Thus, what we actually have is a mixture of the SB and angular
size tests. This is discussed in the next section, where the test is formulated in terms
of some relation (assumed universal) between structural parameters of early type galaxies.
The evolutionary aspects and corrections are discussed in \S 3. The results, based on good
quality data recently published (Pahre et al. 1996; van Dokkum and Franx 1996;
Fasano et al. 1997; Kelson et al 1997; Bender et al 1997) are presented in \S 4.
A discussion of the results together with the conclusions can be found in \S 5. Previous
claims (Pahre et al 1996) that the static case is discarded at the 5$\sigma$ confidence
level are shown to be incorrect. Particular emphasis is made on the almost-lack of
decidability of the test up to $z\sim$1, except perhaps in the K-band.

\section{The scaling relations and the Surface Brightness test}

Kj\ae rgaard et al (1993) have discussed the possibility of using the structural
relations found for early type galaxies to perform the SB test. They considered the SB
intercept of the Fundamental Plane (FP) or Kormendy relation as a mark for the
Tolman test. It's assumed that an intrinsic relation of the form

\begin{equation}
<\mu_f>_0 = A log R_f + B log \sigma + C,
\end{equation}
\noindent
does exist for early type galaxies, where $<\mu_f>$ is the average SB inside the metric
radius R$_f$. The coefficients {\sl A}, {\sl B} and {\sl C} are given by the fit to a
particular set of data, at z$\sim$ 0. At redshift z, the surface brightness will be

\begin{equation}
(\mu_f)_z = (\mu_f)_0 + \Delta_d + \Delta_e + \Delta_K
\end{equation}
\noindent
where $\Delta_d$ is the cosmic dimming (the same for all galaxies at a given redshift),
and $\Delta_e$ and
$\Delta_K$ are the evolution and K-effect (both different, in principle, from galaxy to
galaxy) corrections. To compare data at different redshift, the SB has to be referred to
the same metric size, which implies the transformation of the observed angular sizes,
$\theta_f$, into the corresponding linear sizes.

For the expanding case, the dimming is given by $\Delta_d$ = 10 log(1+z). Assuming that
the evolution and K-corrections at a given z are the same (in magnitudes) for all the
galaxies defining the relation, and that R$_f$ and $\sigma$ do not change with z, the
structural relation will have the same form at any z, $<\mu_f>_z $ =
A log$\theta_f$ + B log$\sigma$ + $(\delta_z)_{ex}$, where
\begin{equation}
(\delta_z)_{ex} = C + 10 log(1+z) + A log(D_{ex}) + \Delta_e + \Delta_K
\end{equation}
\noindent
with (for $\Omega_0 > 0$),
\begin{equation}
D_{ex} = \frac{2c}{H_0}\frac{\Omega_0z+(\Omega_0 - 2)[-1+(\Omega_0z+1)^{1/2}]}
{\Omega_0^2(1+z)^2}
\end{equation}
\noindent
For the Einstein static case, for which there is neither evolution correction (see later) nor
size correction with z,
\begin{equation}
(\delta_z)_{st} = C + 2.5log(1+z) + A log (D_{st}) + \Delta_K
\end{equation}
\noindent
with (see Moles 1991)
\begin{equation}
D_{st} = \frac{c}{H_0}ln(1+z)
\end{equation}
\noindent

The $\delta_z$ term has the same form for any scaling relation involving luminosities and
sizes. Besides the dimming, cosmic evolution (if any), and  K-correction, $\delta_z$
includes another term that depends on the metric (i.e., on the assumed cosmological model)
that is being tested. That term can be neglected when only expanding
models are compared among themselves, but has a sizeable influence when the static case is
also considered. At $z = 0.4$, the difference between the static and expanding
predictions is reduced by about 0.4 magnitudes (for $\Omega_0 =$ 0.2) with respect to
the case of a pure SB test. As we will see in section 4, the conclusions by Pahre et al
(1996) are affected by this problem.

\section{Evolutionary Corrections}

Any test based on the comparison of (assumed) similar objects at different distances has to
deal with all evolutionary effects. Regarding the static case, the stationarity hypothesis
implies that galaxies would be formed at any moment. The same range of evolutionary stages
should be observed in any (large enough) volume, and at any redshift. This is a statement
of statistical nature since it is the population itself that remains the same at any
redshift, not each individual system. No evolutionary corrections have to be applied, and
any structural relation that could be established at z = 0 should be observed at any
redshift: In the static framework, the existence of the FP or the Kormendy relation at
every z is a test in itself.

For the expanding model, the observation of the same structural relation at different z
values implies that the evolution would follow well defined patterns. Unfortunately the
processes of galaxy formation and evolution are still too poorly constrained to explore
the consequences of the fact that the FP or Kormendy relations are well satisfied at
different redshifts. Given that situation, we will consider here that the evolution
only affects the luminosity and colors of the galaxies.

Past claims that elliptical galaxies may have formed through a continuous merging
process lasting till very low red-shift (e.g. Broadhurst et al. 1992) are now seriously
challenged by detailed observations of elliptical galaxies in clusters at
low-to-moderate redshift. The tight correlations between metallicity and velocity
dispersion or color and magnitude (e.g.. Bower et al. 1992; Ellis et al. 1997) strongly
favor the simpler models of {\sl passive luminosity evolution}. Elliptical galaxies
would seem to have formed at red-shift $z_f > 2-3$, with the bulk of star formation
taking place in relatively short ($\sim 0.5-1$ Gyr) periods of time. Since then, their
luminosities would have passively evolved. Though the existence of merging, expected in
hierarchical models for structure formation, is not completely ruled out by the
observations, this would have to happen at high ($z>3$) redshift. Therefore we choose to
consider only passive evolution, and adopted the models by Bruzual \& Charlot (1993; new
version of 1995), with $z_f$ = 4, to compute the K- and evolutionary corrections. We
notice that those models are compatible with other results and interpretations, as they
are able to provide reasonable fits to the counts, as well as to the color and redshift
distribution of galaxies, to the faintest levels (Metcalfe et al. 1996; Pozzetti at al
1996; Campos 1997).

Fortunately, the details of the star formation processes at that remote epoch are expected
to be of little importance for the z range at which the test can be presently performed.
We have explored a decreasing star formation rate (SFR) with e-folding times of 0.3 and 1
Gyr, in combination with two Initial Mass Functions (IMF). The results do confirm that they
have no sizeable influence on the predictions, and the K- and evolutionary corrections are
similar up to z$\sim$1, except in the B-band, where the corrections become strongly
model dependent from z = 0.6 on.

\section{Performing the Tolman test}

The kind of evolution we consider here imposes that the coefficients of the scaling
relation (other than the zero point) should be the same at any redshift. Regarding the FP
relation, we have adopted the values used by van Dokkum \& Franx (1996) for Coma. The
$\delta_z$ values predicted by the different models, for the redshifted $V_z$ band, are
shown in Figure 1. (We use H$_0$ = 50 \kms~ and $\Omega_0$ = 0.2). The expansion prediction
incorporates the corrections for evolution after Bruzual and Charlot (1993) models. It is
clear that the predictions of the static and expanding models are very close till z$\sim$1,
the difference in the cosmic dimming being almost completely compensated by the metric term
and the evolutionary corrections in the expanding case. The same is true for the other
visible photometric bands (see below). For the B-band, the differences between
the different evolutionary models become even larger than those between the static and
expanding cases for $z\ge$ 0.6. Figure 1 shows that data by van Dokkum and Franx (1996),
and  Kelson et al (1997) does fit very well the predictions, as does the data by
Bender et al (1997) for the redshifted B band (not shown here).

We have also considered the Kormendy relation as a tool for the Tolman test. The value
adopted for the coefficient A is 2.77 for the visible bands, and of 2.5 for the K-band, to
be consistent with Pahre et al (1996). As before, the value of A can have some influence
on the predictions, the difference between the static and expansion cases being further
reduced for higher A values. Not surprisingly, the results are similar to those found for
the FP, in the sense that the predictions of the expanding and static models for the
visible photometric bands (shown in figure 2a, and b) are very similar in the considered
redshift
range. The B- and R-band data, taken from Pahre et al (1995, and 1996)  and from Fasano et
al (1997), are remarkably well fitted by the predictions. This is at variance with Pahre et
al, who found that their B-band intercept for A851 was out of the prediction in the expanding
case by 0.3 magnitudes. The difference is probably due to the use of different models to
evaluate the evolutionary corrections.

We therefore conclude that the Tolman test cannot be used to distinguish between the
expanding and non-expanding world models in the optical passbands. The situation looks
however more promising in the K-band, where the expected (model) luminosity evolutionary
corrections
are smaller. Now, the difference between the static and
expanding predictions amounts to $\sim$ 0.4 magnitudes at z = 0.4, and it reaches 0.8
magnitudes at z = 1. Unfortunately, the K-band data is not fitted by any model. The
discrepancy amounts to $\sim$0.2 magnitudes for the expanding case, and to $\sim$0.6
magnitudes for the static case. Trying to understand the situation, we notice that the SB
values at the 1 kpc intercept given for Coma by Pahre et al imply $\mu_B - \mu_K = $ 4.56,
which is significantly redder than expected. Indeed, Persson et al (1979) gave an average
value
$<B-K>$ = 4.22 for a sample of 52 galaxies in Coma, a result  that has been confirmed by more
recent studies (Recillas-Cruz et al 1991), and agrees with the value predicted by Bruzual \&
Charlot models at the redshift of Coma. If charged on the K-band data, this discrepancy could
account for the misfit found for the K-band data and would shift the data points somewhere
between the static and expansion predictions. More data are needed to resolve this issue.

\section{Conclusions}

We have shown that the dependence on cosmology of the Kormendy relation or the FP includes,
besides the SB cosmic dimming and the evolutionary (if any) and K-corrections, a factor
depending on the metric. This metric term, that has been generally overlooked, has a sizeable
effect, when the static predictions are also considered. It explains the
discrepancy between the results reported by Pahre et al (1996) and ours, as they did not take
it into account.

In principle, the SB test could be formulated in such a way to avoid the presence of the
metric term, by using the $\mu_e$-$\sigma$ relation. However, looking at this relation for
the clusters considered here, it is found
that it is too loosely defined, and cannot be used with any confidence for the
Tolman test.

Our main result is that as far as scaling relations are used, the Tolman test cannot be
conclusive in any practicable range of redshift for visible photometric bands data, due
to the conspiracy of the metric term and the evolution corrections to reduce the difference
between the static and expanding world models. In the V- and R- bands those differences are
nearly zero till z$\sim$1. This is also the case for B-band data till z$\sim$0.6. At larger
redshifts the static case is enclosed in the spread of possible evolutionary corrections. The
B-, V-,  and R-band data are shown to fit very well the (similar) predictions of the different
models. Only in the K-band the predictions are separated, even if only by modest amounts till
z$\sim$1, but on the empirical side, the situation is more ambiguous. The data are not fitted
by any prediction, even if they are closer to the expanding case. We have noticed that there
is a clear mismatch between the average $<B-K>$
color index obtained from the 1kpc intercepts of the Kormendy relation, and the measured
values for similar galaxies, or even the predictions at z$\sim$0 by the evolutionary models.
The situation needs to be clarified before a conclusion based on the Tolman test could be
attempted.

The data confirm that the FP and Kormendy relations are well defined at least till z$\sim0.5$.
In the static case they are automatically satisfied since the stationarity hypothesis implies
that there should be always and in every cluster a population of galaxies satisfying a given
relation. In the expanding case it represents only a constraint on the allowed evolutionary
paths of early type galaxies, which should be checked independently of the Tolman test.
It is clear that, if the change of the coefficients of the structural relation with z hinted by
van Dokkum and Franx (1996) were confirmed, the static, non-evolving case would be ruled out. In
that sense, the scaling
relations constitute a test in themselves to prove the evolving character of the world.

{\sl Acknowledgments}. This research was partly supported by DGICYT (Spain) grant PB93-0139,
and the Danish Natural Science Research Council grant 9401635. Carlos Barcel\'o is
acknowledged for helpful discussions on the nature of the tests. We acknowledge the referee
for enlightening comments that did help to appreciably improve the manuscript.

\clearpage

\figcaption{Predictions for $\delta_z$ from the FP relation in the $V_z$ band, for
the static
(dashed line) and expanding models (solid lines). For these we took z = 4 as the epoch of
galaxy formation, and considered an exponentially decreasing star formation rate with
e-folding time 0.3 and 1 Gyr in combination with two different IMFs, the Scalo and Salpeter
ones. Only the evolution corrections, evaluated with Bruzual and Charlot models were
considered as the data are free from the K-effect. The data shown corresponds to the zero
points of the FP relation found by van Dokkum and Franx (1996; filled squares), and by
Kelson et al (1997; circles).}

\figcaption{Predictions for $\delta_z$ from the Kormendy relation in the B-, R- and K-
bands, for the static (dashed line), and expanding (solid lines) models. The expanding
models include the evolutionary and K corrections from Bruzual \& Charlot models. The
B-band data are from Pahre et al (1996). The R-band data are from Pahre et al (1996;
filled squares), and from Fasano et al (1997; circles). Data for the K-band are from Pahre
et al (1996). The errors reported by the different authors are not larger than the size of
the symbols. }

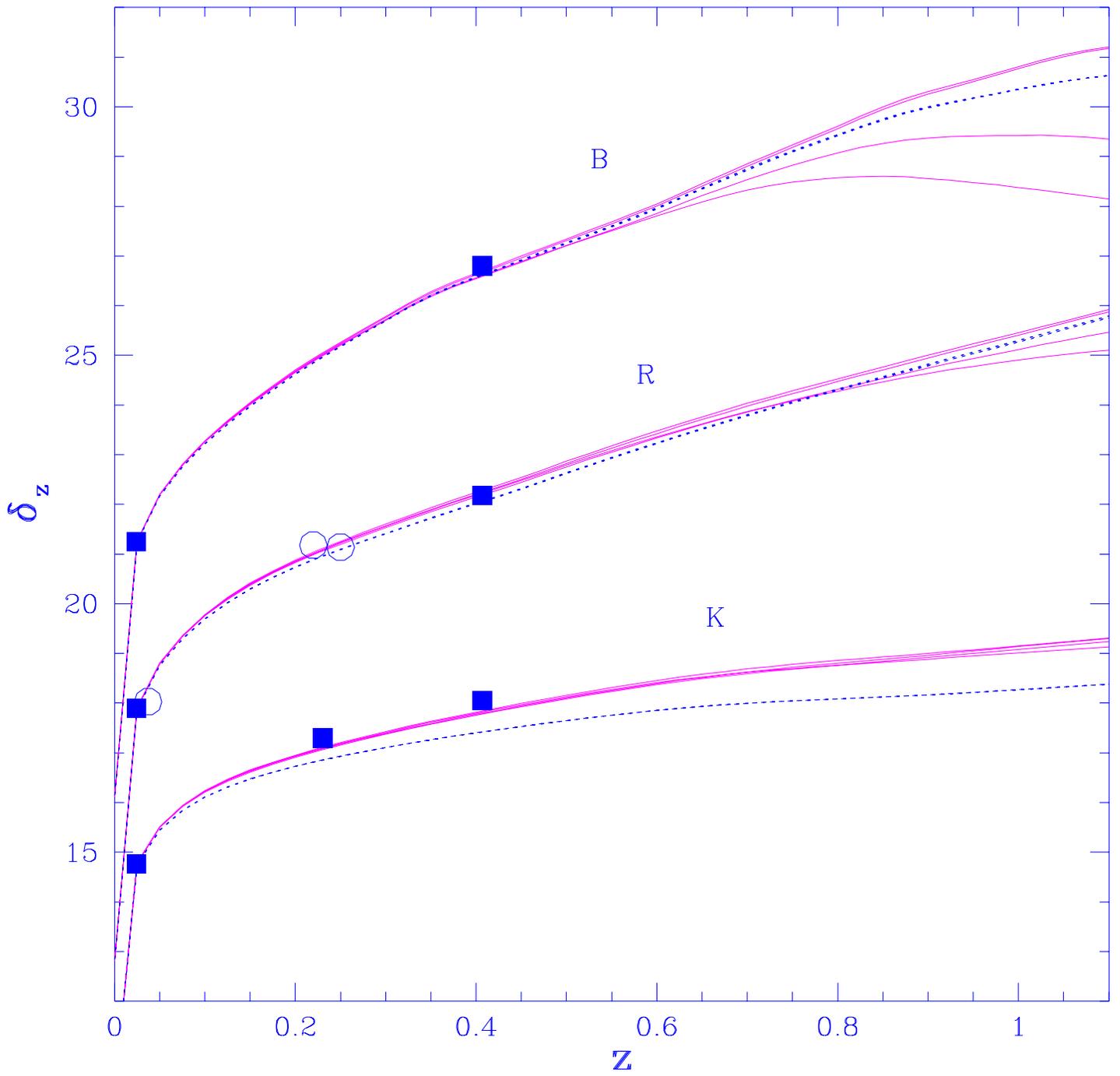
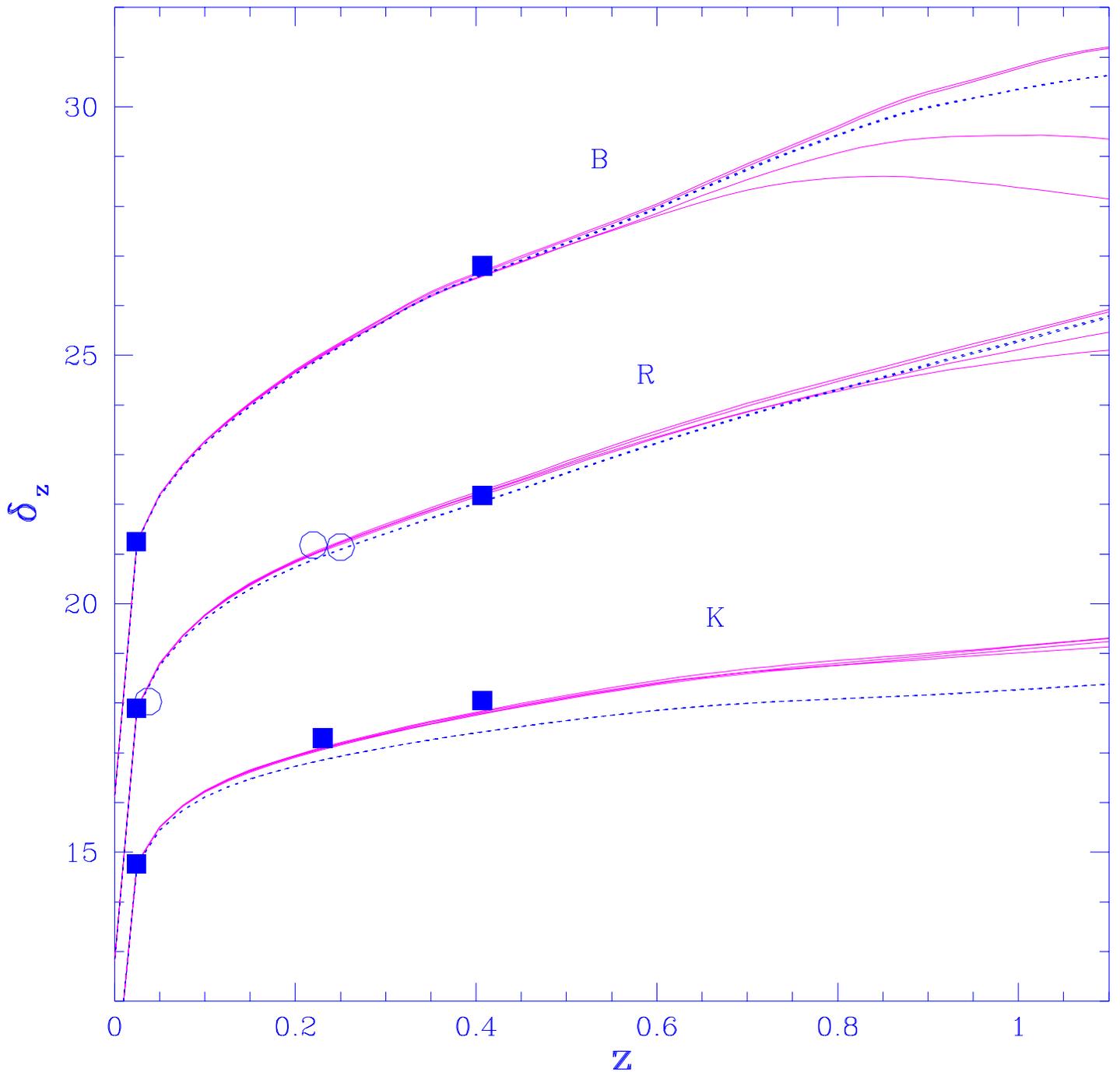
\end{document}